\documentclass[pra,twocolumn]{revtex4} 
\usepackage{graphicx} 
\begin{document}

\bibliographystyle{unsrt}

\title{Demonstration of Non-Deterministic Quantum
 Logic Operations \\using Linear Optical Elements} 
\author{T.B. Pittman, B.C. Jacobs, and J.D. Franson} 
\affiliation{The Johns Hopkins University, 
Applied Physics Laboratory, Laurel, MD 20723}

\date{\today}

\begin{abstract} 
Knill, Laflamme, and Milburn recently showed that
non-deterministic quantum logic operations could be performed using linear
optical elements, additional photons (ancilla), and post-selection based on the
output of single-photon detectors [{\em Nature} {\bf 409}, 46 (2001)].  Here we
report the experimental demonstration of two logic devices of this kind, a
destructive controlled-NOT ({\scriptsize CNOT}) gate and a quantum
parity check.  These two devices can be combined with a pair of
entangled photons to implement a conventional (non-destructive)
 {\scriptsize CNOT} that succeeds with a
probability of $\frac{1}{4}$.
\end{abstract}

\maketitle

One of the main difficulties in any optical approach to quantum information
processing is that nonlinear interactions between single photons are requried
to implement logic devices that operate with 100\% efficiency.  Recently,
however, Knill, Laflamme, and Milburn (KLM) \cite{knill01} showed that
probabilistic quantum logic operations could be performed using only linear
optical elements, additional photons (ancilla), and post-selection based on the
output of single-photon detectors.  These devices succeed in producing the
desired logical output with a probability that can approach unity and they may
form the basis for a scalable approach to quantum computing.  

Here we report the
experimental demonstration of two logic devices of this kind,
a destructive controlled-{\scriptsize NOT} 
({\scriptsize CNOT}) gate and a quantum parity
check \cite{pittman01}. The output of the  destructive {\scriptsize CNOT} 
({\scriptsize DCNOT})  gate is identical to
that of a conventional (non-destructive)  {\scriptsize CNOT} 
gate except that the
information associated with the control photon is  destroyed during
the operation.
However, the destructive {\scriptsize CNOT} gate and quantum parity check 
demonstrated here can be combined with a pair of entangled photons
to implement a conventional {\scriptsize CNOT} gate
that succeeds with a probability of  $\frac{1}{4}$ \cite{pittman01}.

In the original approach suggested by KLM \cite{knill01}, the value of
each qubit is represented by a single photon that may be located in 
one of two paths, such as
two optical fibers.  One path represents a logical value of 0 while the other
path represents a value of 1.  The logic operations involve interference between
different optical paths, which can be very sensitive to thermally-induced phase
shifts.  In contrast, we use polarization-encoded qubits \cite{bouwmeesterbook}
in which the values of the qubits are represented by the two linear states of 
polarization of a single
photon, with a horizontal polarization state $|H\rangle$ representing the value
0 and a vertical polarization state  $|V\rangle$ representing the value 1.  
The use of polarization-encoded qubits can eliminate the need for any interference between
two different optical paths \cite{koashi01}, which may be an advantage in
practical applications.

In an earlier theory paper \cite{pittman01}, we showed 
that polarization-encoded qubits and post-selection could
 be used to implement a number of non-deterministic quantum logic
operations, including the quantum parity check and destructive {\scriptsize
CNOT} gate shown in Figure 1. 
\begin{figure}[b]
\includegraphics[width=2.75in]{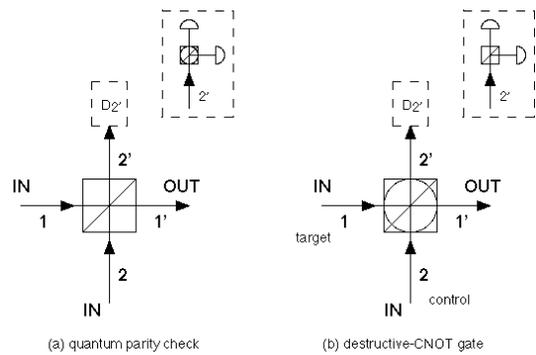} 
\caption{Implementation of a quantum parity check (a) and a 
destructive {\scriptsize CNOT}
gate (b) using polarizing beam splitters.  The polarizing beam splitter (PBS)
is oriented in the HV computational basis for the parity check and rotated by
$45^{o}$ for the destructive {\scriptsize CNOT}.  
The polarization-sensitive detector $D_{2'}$  consists of a
second polarizing beam splitter and two single-photon detectors.  It is oriented in the
HV basis for the destructive {\scriptsize CNOT} and rotated by $45^{o}$ for the parity
check.}  
\label{fig1} 
\end{figure}
  In each of these devices, two input qubits in the form of two single
photons are incident on a polarizing beam splitter that completely transmits
horizontally-polarized photons and totally reflects vertically-polarized photons.
 A polarization-sensitive detector in output path 2' detects any photons in
either the horizontal-vertical (HV) computational basis or another basis
rotated by $45^{o}$.  As shown in the insets, 
the polarization-sensitive detectors can be implemented
using another polarizing beam splitter and two single-photon detectors.  The
output of the devices is accepted only for those cases in which the detector
registers one and only one photon.  This post-selection process provides the
origin of the nonlinearity required for the logic operations, since the detection
process is inherently nonlinear.

Although the physical stucture of these two devices is  similar, the
operation and desired output 
of each is very different. The operation of the quantum parity check 
  is the
most straightforward and will  be described first, although the demonstration
of the destructive {\scriptsize CNOT} gate is the most important result
presented here.
The goal of the quantum parity check shown in
Figure 1a is to transfer the value of the input qubit in path 1 to the output in
path 1' provided that its value is the same as that of the second input qubit
in path 2.  If the qubits have different values, the device fails and produces no
output.  The parity-check operation can be understood from the basic properties
of a polarizing beam splitter\cite{pan01a}.   If only one photon is to be
detected, both of the incident photons must be transmitted or
 both must be reflected.  In either
case, the polarizations and the corresponding values of the qubits must be the
same.  This ability to compare the
polarizations of two photons has been proposed for use  
in  entanglement purification \cite{pan01a}
and a variety of other quantum information processing applications
 \cite{pan98,bouwmeester99,bouwmeester01,yamamoto01}.

The parity-check operation described above is essentially classical in nature,
 but quantum-computing applications require that it be
performed without measuring or determining the values of either qubit.
For the probabilistic quantum parity check illustrated in Figure 1a, this is
accomplished by orienting the polarization-sensitive detector in a
basis rotated by $45^{o}$, which essentially erases any
information regarding the value of either of the input qubits.  When the inputs
consist of an arbitrary superposition of states,
this quantum erasure \cite{scully82,kwiat92} technique combined with the
post-selection process 
 maintains the required coherence of the probability amplitudes, 
 as will be
 described below. 
 
 The implementation of the destructive {\scriptsize CNOT} gate
  shown in Figure 1b is similar to the
 quantum parity check,  except that the polarizing beam splitter is
 oriented in a basis rotated by $45^{o}$  and the
polarization-sensitive detector is oriented in the HV basis.  The
goal of the  destructive {\scriptsize CNOT}
 gate is to flip the logical value of the target qubit 
(eg. $0 \leftrightarrow 1$) if the control qubit has the value 1 and do nothing if the
control  qubit has the value 0.  The expected output of the gate shown in
 Figure 1b is that of a conventional {\scriptsize CNOT}, except that
the information contained in the control qubit is destroyed by the
polarization-sensitive detector. 
The operation of the destructive {\scriptsize CNOT}
gate can be understood by expanding the input qubit polarization states
in the $45^{o}$ basis of the polarizing
beam splitter and then re-expressing the output states in the HV basis of the
detector.  It can be shown \cite{pittman01} that an
arbitrary target state will be flipped if the control photon in mode 2 was
vertically polarized (logical value 1) but it will be unaltered for a
horizontally polarized control photon (logical value 0):

\begin{eqnarray}
Control = 1: \hspace{.15in}
\alpha|H\rangle_{1}+\beta|V\rangle_{1} \rightarrow
 \alpha|V\rangle_{1'}+\beta|H\rangle_{1'}
 \nonumber\\
Control = 0: \hspace{.15in}
\alpha|H\rangle_{1}+\beta|V\rangle_{1} \rightarrow
 \alpha|H\rangle_{1'}+\beta|V\rangle_{1'}
\label{eq1}
\end{eqnarray}

 Because of the similarity of these two devices, we were able to
demonstrate both logic operations using a single experimental set-up, which is
shown schematically in Figure 2.  Here the polarizing beam splitters of Figures
1a and 1b correspond to the second polarizing beam splitter, PBS2, in Figure 2. 
The initial polarizing beam splitter, PBS1, was simply used to separate the two
photons produced in a type-II parametric down-conversion \cite{shih94,rubin94}
crystal (BBO),
 which produces pairs of photons, one polarized vertically, the other polarized
horizontally.  The two photons were then directed along two different paths
towards PBS2.  The two paths were adjusted to the same length using mirrors on
translation stages, which was necessary in order to ensure that both photons
arrived at the second beam splitter at the same time\cite{pittman96}.  The
polarization of either photon could be rotated into any desired orientation using half-wave
birefringent plates located in each path. 

\begin{figure}[b]
\includegraphics[width=3in]{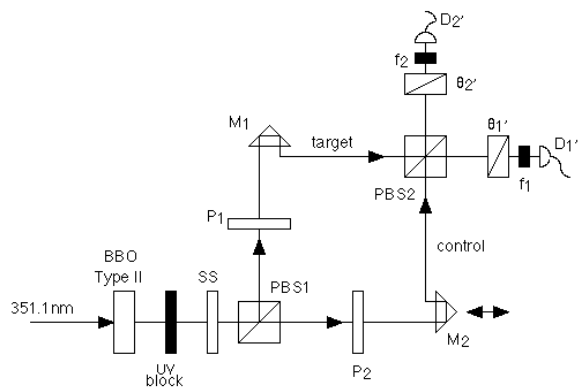} 
\caption{ A simplified outline of the experimental apparatus used to demonstrate the
quantum parity check and destructive {\scriptsize CNOT} operations of Figure 1. 
 A 1.0 mm thick 
BBO down-conversion crystal (type-II) was pumped by roughly 30 mW of the 351.1 nm
line of an Argon-Ion laser.  PBS1 and PBS2 were
polarizing beam splitters while $M_{1}$ and $M_{2}$ were
mirrors mounted on translation stages.  Half-wave plates $P_{1}$ and $P_{2}$ 
were located in each path while $\theta_{1'}$ and $\theta_{2'}$ were
polarization analyzers in front of detectors $D_{1'}$ and $D_{2'}$. $f_{1}$ and
$f_{2}$ were 10 nm FWHM bandpass filters centered at 702 nm, and
 SS was a birefringent element for Shih-Sergienko compensation
\protect{\cite{shih94}.}}
\label{fig2}  
\end{figure}

The polarization-sensitive detectors of
Figures 1a and 1b were implemented using a rotatable polarization analyzer 
$\theta_{2'}$  and
one single-photon detector in output path 2'.  In contrast to the
polarization-sensitive detectors of Figure 1, this arrangement could only detect
one of the possible output polarization states and reduced the probability of
success from $\frac{1}{2}$ to  $\frac{1}{4}$ \cite{pittman01}.  
Another polarization analyzer $\theta_{1'}$  and a second
single-photon detector were used to measure the polarization state of the photons in the
logical output path 1', which allowed a comparison between the expected and
actual outputs of the devices. 
Since only two photons were incident at any
given time, the post-selection process described above was equivalent to
monitoring the rate of  coincident detection events as a function of various
combinations of wave plate and analyzer orientations.

 The experimental results
from the quantum parity check operation are summarized in Figures 3 and 4.  The logical
truth table of Figure 3 shows the relative probability of obtaining an output
value of 0 or 1 for all of the possible combinations of
input values.  The half-wave plates were oriented 
in this case to give input photons with horizontal
or vertical polarizations (0 or 1) and the polarization analyzer $\theta_{1'}$ 
was oriented horizontally or vertically.  It can be seen that the results obtained agree with
what would be expected from a parity check to within an error on the order of
1\%.

\begin{figure}[b]
\includegraphics[width=3in]{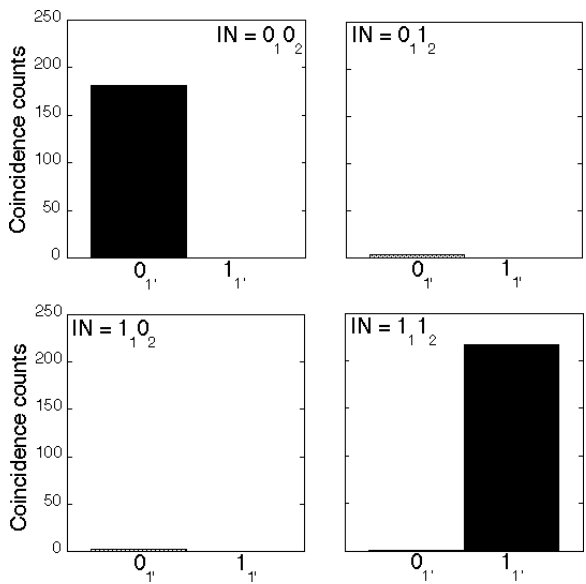} 
\caption{ Logical truth table summarizing the results from the
quantum parity check.  The
number of coincidence events per minute corresponding to 0 or 1 that were
obtained for all possible values of the input qubits is shown.  The results are
consistent with what would be expected from a parity check operation.} 
\label{fig3} 
\end{figure}

 A more
intriguing property of the quantum parity check is illustrated in Figure 4,
 where the wave plates were adjusted so
that input 2 was in the state 
$|in\rangle_{2}=\frac{1}{\sqrt{2}}(|0\rangle +|1\rangle)$.
 As we showed in Ref. \cite{pittman01}, an arbitrary input
state in path 1 is expected to be coherently transferred into output path 1'
for this choice of input 2.  It is interesting to note that the polarizing beam
splitter PBS2 transmits the horizontal component of input 1 into output path 1'
as required, but it totally reflects the vertical component into detector
$D_{2'}$,
where it is destroyed.  As a result, the device must somehow replace the
arbitrary vertical component from input 1 with a vertical component from input 2
(which is reflected into output path 1') even though that component initially
had a fixed value of $\frac{1}{\sqrt{2}}$.  In essence, the post-selection
process makes the polarizing beam splitter transparent to all components of 
input 1.
   For the data shown in Figure 4, input 1 was arbitrarily chosen to be in the state
$|in\rangle_{1}=0.94|0\rangle +0.34|1\rangle$,
 which corresponds to a photon polarized at an angle
of approximately $20^{o}$.   It can be seen that the output state corresponds
to the same polarization to within the experimental error,
 which demonstrates the coherent
nature of the quantum parity check operation.

\begin{figure} 
\includegraphics[width=2.5in]{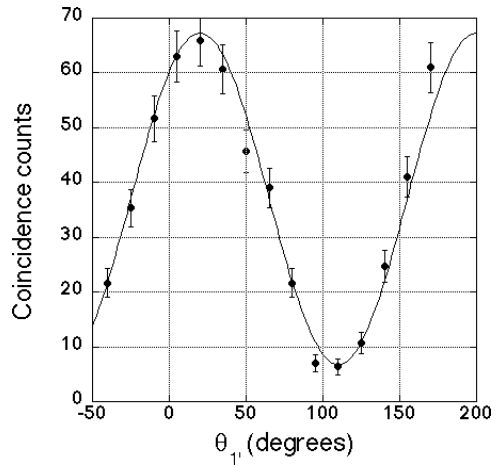} 
\caption{ Experimental test of the coherence of the output of the 
quantum parity check. 
The number of coincidence events per minute is plotted as a function of the
setting of polarization analyzer $\theta_{1'}$  for an input 1 state
corresponding to a polarization of $20^{o}$. 
 The solid curve represents an empirical fit to the data. 
The measurements are consistent with an output state with the same polarization,
which demonstrates the coherence of the parity check when acting on superposition
states.} 
\label{fig4} 
\end{figure}

The main result of this Letter is shown in Figure 5, which shows a 
logical truth table summarizing the
experimental results for the destructive {\scriptsize CNOT} gate. 
Rather than physically rotating the polarizing beam splitter through a $45^{o}$
angle as shown in Figure 1b, it was more convenient to 
rotate the incident photons and the detector bases
using the half-wave plates.  As mentioned above, the intended function of the
destructive {\scriptsize CNOT} gate is to flip the target
 bit if and only if the control bit is equal
to 1.  It can be seen from Figure 5 that our destructive {\scriptsize CNOT}
 gate performs this operation
with an average error of approximately 17\% for the case where all of the input
qubits have the value 0 or 1.  There are superposition states, however, for which
the errors are much less than 1\%, as is the case for the parity check data of
Figure 3.  The mean error for the current set-up when averaged over all possible
input states is approximately 8\% as a result.  These errors were largely due to the optical
quality of the polarizing beam splitters, which were of commercial grade and had
distortions on the order of $\frac{1}{4}$-wave.  Much lower error rates could
be obtained using custom-made polarizing beam splitters and single mode optical fibers.  In
any event, the controlled target state-flipping operation of the 
destructive {\scriptsize CNOT} is
clearly evident from the data shown in Figure 5.

\begin{figure}
\includegraphics[width=3in]{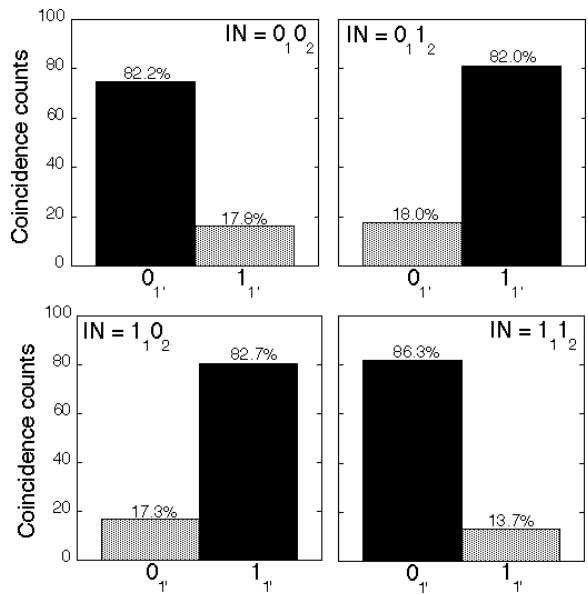}
\caption{Logical truth table summarizing the results obtained from the
destructive {\scriptsize CNOT}.  The number of coincidence events per minute corresponding
to 0 or 1 that were obtained for all possible values of the input qubits is shown. 
The results show that the destructive {\scriptsize CNOT} flips the value of the target qubit
when the control qubit has a value of 1, as desired.  The errors shown here
correspond to the worst case and could be greatly reduced using custom-made
optics or optical fibers.} 
\label{fig5} 
\end{figure}

The destructive {\scriptsize CNOT} operation was also tested using
superpositions of input states.  Coherent results similar to those of Figure 4
were obtained but are not included here.  Once again, this shows that the 
operation was performed without measuring the values of the
qubits\cite{pittman01}.

 The results presented here demonstrate that certain non-deterministic quantum logic
operations can be performed in a straightforward manner using linear optical
elements \cite{knill01}.   As we have shown earlier
\cite{pittman01},
 the quantum parity check
 and destructive {\scriptsize CNOT} described above can be combined
 with a pair of entangled photons to implement a non-destructive {\scriptsize CNOT}
  operation with a
probability of success of $\frac{1}{4}$.  Experiments of that kind will require
the production of four photons in the same spatial and temporal mode 
\cite{zukowski95,zeilinger97,ou97}, whereas
our current experiments required only two photons.  Zeilinger's group has
recently
demonstrated \cite{pan01b} a
method for producing the necessary four-photon state, so that the
demonstration of a full non-deterministic {\scriptsize CNOT}
 gate should be feasible.  A more difficult
challenge for practical applications is the requirement that the probability of
success be on the order of unity as is required by the current state of quantum
error correction \cite{nielsenchuang}.
  In principle, KLM \cite{knill01} have shown
that it is possible to use additional ancilla and detectors
 to achieve a success rate arbitrarily
close to one.  We expect that further work in that area will be required in order
to develop a practical approach to quantum computing.  In any event, the
experimental results presented here may provide a first step towards that goal.

This work was
supported by the U.S. Office of Naval Research and by Independent Research and
Development funds.  We would like to acknowledge valuable discussions with M.
Donegan and M. Fitch.



\end{document}